\documentclass[a4paper,12pt]{article}
\pdfoutput=1
\usepackage{amsfonts,amsthm,amsmath,amssymb,graphicx,hyperref}
\usepackage[vcentermath]{youngtab}
\usepackage{ytableau}

\def\bra#1{\left\langle #1\right|}
\def\ket#1{\left| #1\right\rangle}

\newcommand{\tr}{\operatorname{tr}}

\def\Tr{{\rm Tr\, }}

\newcommand{\be}{\begin{equation}}
\newcommand{\bea}{\begin{eqnarray}}
\newcommand{\ee}{\end{equation}}
\newcommand{\eea}{\end{eqnarray}}

\begin{document} 

{\LARGE{\centerline{\bf Gauged fermionic matrix quantum mechanics}}}  

\vskip.5cm 

\thispagestyle{empty} 
\centerline{ {\large\bf David Berenstein$^a$\footnote{{\tt dberens@physics.ucsb.edu}} 
and Robert de Mello Koch$^{b,c,}$\footnote{{\tt robert@neo.phys.wits.ac.za}}}}

\vspace{.4cm}
\centerline{{\it ${}^{a}$ Department of Physics, University of California,}}
\centerline{{\it Santa Barbara, CA 93106, USA}}

\vspace{.4cm}
\centerline{{\it ${}^{b}$ School of Physics and Telecommunication Engineering},}
\centerline{{ \it South China Normal University, Guangzhou 510006, China}}

\vspace{.4cm}
\centerline{{\it ${}^{c}$ National Institute for Theoretical Physics,}}
\centerline{{\it School of Physics and Mandelstam Institute for Theoretical Physics,}}
\centerline{{\it University of the Witwatersrand, Wits, 2050, } }
\centerline{{\it South Africa } }

\vspace{1truecm}

\thispagestyle{empty}

\centerline{\bf ABSTRACT}

\vskip.2cm 

We consider the gauged free fermionic matrix model, for a single fermionic matrix. In the large $N$ limit this system describes a $c=1/2$ chiral fermion
in $1+1$ dimensions.
The Gauss' law constraint implies that to obtain a physical state, indices of the fermionic matrices must be fully contracted,
to form a singlet.
There are two ways in which this can be achieved: one can consider a trace basis formed from products of traces
of fermionic matrices or one can consider a Schur function basis, labeled by Young diagrams.
The Schur polynomials for the fermions involve a twisted character, as a consequence of Fermi statistics. 
The main result of this paper is a proof that the trace and Schur bases coincide up to a simple normalization coefficient that we have computed.

\setcounter{page}{0}
\setcounter{tocdepth}{2}
\newpage
\tableofcontents
\setcounter{footnote}{0}
\linespread{1.1}
\parskip 4pt

{}~
{}~

\section{Introduction}

The discovery of the gauge/gravity duality \cite{Maldacena:1997re} has made precise how a theory of physical strings in higher dimensions arises from the large $N$ limit of gauge theories, as originally suggested by 't Hooft \cite{tHooft:1973alw}. Such a string theory (with appropriate boundary conditions) is in fact now considered to be equivalent to gauge theory. This progress, in principle, has given us a fully consistent description of quantum gravity. 

Unfortunately, many very interesting and difficult problems in quantum gravity are still largely inaccessible in the dual gauge theory dynamics: the dual gauge theory needs to be solved in the strong coupling regime. 

It is often useful to study simpler, even exactly solvable,  large $N$ models in order to try to understand better the emergence of the additional dimensions in string theory. 
Many times such models arise as (possibly protected) sectors of a larger theory, or they can be a starting point to do perturbation theory in some coupling constant. 

This strategy has been very successful in the case of the half BPS sector in ${\cal N}=4$ SYM theory. This sector is actually protected by supersymmetry and is generated by traces of a single scalar field $\tr(Z^k)$, where $Z$ is a highest weight state for $SO(6)$ in the ${\cal N}= 4$ gauge multiplet.
The first important result was the full exact diagonalization of the two point function \cite{Corley:2001zk}. This was done using combinatorial techniques to express a complete basis of states in terms of of Young diagrams. The naive basis of traces is not orthogonal: there are non-trivial overlaps generated at order $1/N$. 
These overlaps generate  complications when analyzing  anomalous dimensions in other setups.

The main statement of that paper is that the states built this way are orthogonal.
It was then realized that such a system could be thought of in terms of a 2D fermion description, similar to the integer quantum hall effect \cite{Berenstein:2004kk}. 
This description it made clear that strings could be interpreted as chiral edge excitations of a droplet, and also that it is possible to identify a class of D-branes (giant gravitons and dual giant gravitons \cite{McGreevy:2000cw,Grisaru:2000zn,Hashimoto:2000zp}, see also \cite{Balasubramanian:2001nh}) in terms of single fermion and hole states in the droplet dynamics.

A big surprise  is that the fermion droplet prescription also describes the supergravity solutions exactly \cite{Lin:2004nb} and for each droplet configuration in the gauge theory one can find a solution of supergravity that describes it. This sector alone has led to numerous additional insights in the theory of quantum gravity. 
Making the combinatorial problem of relating the traces and Schur functions more precise,  it has been noted that the topology of spacetime can be changed by superposing states of a fixed topology \cite{Berenstein:2016pcx,Berenstein:2017abm}. 

In this paper we study the fermion counterpart of this dynamics: a single fermionic gauged matrix model. The states arising here can be thought of as a special class of states of the $SU(1|1)$ sector of $N=4$ SYM.  These states are built from products of traces of a single Weyl fermion (spin up) with $\tr(\psi_+^k)$, but they are not protected by symmetry: a non-trivial anomalous dimension is generated at higher loop orders in perturbation theory. This sector has been analyzed in some detail in \cite{Staudacher:2004tk}.

Just like in the half BPS sector, the fermionic matrix model can be studied in its own right. There is a basis of traces and another basis based on Young diagrams (Schur functions). This paper studies in detail the relationship between these two and we find that surprisingly, they are the same basis, although they have different normalizations.
In contrast to the bosonic case, the Schur functions for the fermions involve a twist due to Fermi statistics. This has consequences for the map between basis states, which 
produce non-trivial factors that are square roots of integers. These arise when the representations of the symmetric group are treated explicitly using the Young orthogonal representation. 

The rest of the paper is organized as follows.
In the next section we introduce the gauged fermionic matrix model and review relevant background from the
corresponding bosonic model. This section also develops a precise statement of the conjectured relation between the trace and Schur function bases. Section \ref{fermionicschurs} reviews the construction of the Schur functions for fermions. The novel ingredient in the construction is a twisted character, considered in detail in section \ref{twistedchar}. This discussion is enough to prove a special case of the general conjecture. In section \ref{proof} we give a complete proof of our conjecture. The proof uses elements from the representation theory of both the symmetric and unitary groups, combinatorics, orthogonality at infinite $N$ and the ring structure of multiplying by traces, to develop an induction argument. We draw conclusions and suggest some avenues for further study in section \ref{conclusions}.

\section{Preliminaries}\label{preliminaries}

The gauged fermionic matrix model is defined by the following first order action
\begin{equation}
S=\int dt \tr\left[\bar \psi i D_t \psi - m \bar \psi \psi\right]
\end{equation}
where $D$ is a covariant derivative and $\psi$ transforms in the adjoint of $U(N)$. 
If we choose the gauge $A_0=0$ the dynamics is free with $\psi$ acting as raising operators and $\bar \psi$ as lowering operators giving rise to a fermionic Fock space of states.

The vacuum is gauge invariant (a singlet of $U(N)$). Any state in the dynamics can be accessed by raising operators acting on the vacuum. However, we need to impose the Gauss' law constraint. A fermion $\psi^a_b$ has an upper and lower $U(N)$  index. These need to be fully contracted  to form a singlet.
These contractions are in the form of traces. For example, a single trace state acting on the vacuum is given by
\begin{equation}
\ket k = \tr(\psi^k)\ket 0
\end{equation}
Using the cyclic property of the trace and the fermionic character of $\psi$ it can be easily shown that
\begin{equation}
\tr(\psi^k)= (-1)^{k-1}\tr(\psi^k)
\end{equation}
so that only traces with an odd number of fields $\psi$ are allowed. Each of these traces has fermi statistics and has energy $k=2s+1$, where $s=0,1, \dots$.
At infinite $N$ each trace is supposed to correspond to a different `particle'.

If we normalize the energies in units of $1/2$, we get a single particle state for each half integer $(s+1/2)$. This is the same spectrum of states as a free chiral fermion in $1+1$ dimensions (a $c=1/2$ CFT) on a circle with NS boundary conditions  \cite{Berenstein:2004hw} (antiperiodic boundary conditions $\chi(\theta+2\pi)=-\chi(\theta)$).

This suggests that this is a matrix model for a $c=1/2$ CFT in a similar way that a bosonic gauged matrix model gives rise to a $c=1$ chiral  boson in $1+1$ dimensions. 
The main insight of this map to the chiral boson is that in the bosonic matrix model one can reduce the dynamics to the eigenvalues of the matrix $X$. When the Gauss' law is implemented the eigenvalues act as fermions. This dynamics for bosons is usually best described in terms of a first order dynamics
\begin{equation}
S=\int dt\tr\left[\bar Zi D_t Z- m \bar Z Z \right]
\end{equation}
where $Z= X+iP$ and $\bar Z$ are complex and the Gauss' law constraint requires them to commute. The corresponding fermions are in the phase space of $X$ and the ground state can be described 
by a droplet in phase space. This is a familiar story of the quantum hall effect. The eigenvalue dynamics is the effective field theory of free electrons in the lowest Landau level in $2+1$ dimensions, which has been slightly deformed by a binding potential proportional to  $x^2+y^2$. The traces $\tr(Z^k)$ are collective excitations of the droplet with angular momentum $k$ on the edge.

Unlike the bosonic matrix model, in the fermionic setup we can not choose a gauge where we diagonalize the fermionic degrees of freedom. The interpretation of an edge dynamics is harder to do and will not be pursued here.

Because the bosonic system can also be interpreted in terms of first quantized fermions in a bulk $2+1$  system, we can also write the wave functions in terms of Slater determinants of single particle states. These wave functions are governed by Schur polynomials of the matrices. These are obtained by traces in irreducible representations of $U(N)$, which are labeled by Young diagrams.
The map from multi-traces to Schur functions in the bosonic model is non-trivial \cite{Stone:1990ir}. It is also obtained from character expansions of the symmetric group.
Edges of strongly coupled quantum systems can exhibit $c=1/2$ Majorana modes (see for example \cite{Kitaev:2006lla,Kit2}).

One of the main goals of this paper is to write the corresponding map from traces to Schur functions for the fermionic matrix model. What we will obtain is that the two basis of states, the one of traces and the one of fermionic Schur functions, are actually the same basis up to normalization. In this sense the fermionic matrix model is in the end simpler than the bosonic counterpart.

Let us describe this a little more carefully. 
The fermion dynamics is actually free, and the only constraint is the $U(N)$ invariance. A free system actually has a larger symmetry $U_L(N)\times U_R(N)$ where the first $U_L(N)$ rotates only the upper indices of the fermions (as a fundamental) and the second $U_R(N)$ rotates only the lower indices (as an antifundamental).
If we take a $k$ particle state in the Fock space, it will have $k$ upper indices and $k$ lower indices. We want to decompose the state into irreducible representation of $U_L(N)$. 
Since the state is in a tensor product of $k$ fundamentals, it will be decomposed into irreducible representations that are labeled by a Young diagram: one needs to symmetrize or antisymmetrize the tensor indices. 
These are distinct irreducibles, they have different Casimir's and because the action of $U_L(N)$ is unitary, any pair of states labeled by two different diagrams are  orthogonal to each other. The same analysis can be done with the lower indices. Fermi statistics guarantees that symmetrizing in the upper indices corresponds to antisymmetrizing on lower indices. 
Now, we need to gauge the diagonal embedding $U(N)\to U(N)_L\times U_R(N)$. We thus need the Young diagram representation of the upper indices to be the same as the one for the lower indices, but these are mirrors of each other. 
Thus only diagrams that are self-conjugate are allowed.
There is a unique singlet for such self-conjugate representation. This will be called the Schur function. To properly define the Schur function requires building the map more carefully, which will be described in the next section.

Now, we want to motivate how the Schur functions for fermions and the traces basis should be the same.

To motivate this equality, consider the following identity for the bosonic matrix model
\begin{equation}
\tr(Z^k) = \sum_{single\ hooks} (-1)^{s-1} \begin{ytableau}
\ & \dots &\ \\
\vdots\\
\ \end{ytableau}\label{eq:sihook}
\end{equation}
where the sum is over Young diagrams given by  hooks with $k$ boxes, and $s$ is the number of rows of the diagram.
A precise accounting of how traces act on the full Schur basis can be found in \cite{Berenstein:2017abm}. Acting with a trace adds skew hooks of length $k$ to a given Young diagram in all possible ways with a sign that is $-1$ if the skew hook is extended over an even number of rows.

If we {\em naively} substitute fermions in equation \eqref{eq:sihook}, basically arguing by analogy, we get quite a few restrictions because the only allowed Young tableaux are those that are equal to their reflection about the diagonal.
Basically, if a similar equation holds for fermions, there is only one hook that appears in the sum. In this sense, a natural guess is that each trace is equal to a hook
\begin{equation}
\tr(\psi^{2s+1}) =  \alpha_s \begin{ytableau}
\ & \dots &\ \\
\vdots\\
\ \end{ytableau}
\end{equation}
where the single row and single column each has $s+1$ boxes. Here we allow the possibility of a non-trivial normalization factor. The origin of this factor is that at leading order in $N$ we have that
\begin{equation}
\bra 0\tr(\bar\psi^{2s+1})\tr(\psi^{2s+1})\ket 0 \simeq (2s+1) N^{2s+1}
\end{equation}
whereas for Young tableaux with $k$ boxes we usually have that
\begin{equation}
\bra 0 \overline{YT}(k) YT(k)\ket 0 \simeq N^k
\end{equation}
In these equations we use fields with canonical normalization. Other normalizations are possible so that the right hand side has no powers of $N$, and these are useful for taking the strict $N\to \infty$ limit.

This suggests that $\alpha_s= \sqrt{2s+1}$. This is very different to the bosonic matrix model where the coefficients in the translation are all $\pm 1$ and the factor of $2s+1$ just comes from the number of Young diagrams that contribute. 

Also, if we consider the general action of the product of a trace on a given Schur function, which is by adding skew hooks of length $2s+1$ in all possible ways, the condition of reflection symmetry of the allowed Young tableaux
means that there is only one place where the skew hook can be attached: it must be attached symmetrically with respect to the diagonal. The hook can only be attached if the diagram does not already contain a hook of the given length. This is indicative of the Fermi statistics of the traces where one does not allow double occupation of a state.

Now, let us describe the conjecture we will prove in this paper. Consider a trace structure 
\begin{equation}
\tr(\psi^{2s_1+1}) \dots \tr(\psi^{2s_k+1})\ket 0
\end{equation}
describing a state in the field theory. 
We will show that this state is (up to a normalization coefficient) equal to the state given by the tableau
\begin{equation}
\begin{ytableau}  {\bf 1}& 1& 1& \dots & 1\\
1& {\bf 2 }& \dots &2 \\
1& \vdots &\ddots\\
\vdots&2 \\
1
\end{ytableau}
\end{equation}
where there are exactly $2s_1+1$ ones (the largest hook on the diagonal has $2s_1+1$ boxes), there are exactly $2s_2+1$ boxes with a label $2$ etc, where the label just indicates how we associate different traces to different hooks. 
The bold-face numbers are on the diagonal and they label the hooks. 
We will also show that the normalization coefficient is $\pm \prod_i \sqrt{2s_i+1}$.

\section{Schur polynomials for fermions}\label{fermionicschurs}

The Schur polynomial basis constructed in \cite{Corley:2001zk} for a single adjoint scalar, diagonalizes the free field
two point function and manifestly accounts for the trace relations that appear at finite $N$.
In this section we will review the analogous construction, for a single adjoint fermion, given in \cite{Koch:2012sf}.

Consider a single fermion $\psi^i_j$ transforming in the adjoint of the gauge group $U(N)$.
The two point function is
\bea
 \langle \psi^i_j (\psi^\dagger)^k_l\rangle = \delta^i_l \delta^k_j
\eea
Since fermionic fields anticommute, it is important to spell out how products of the fermion fields are ordered.
With the convention
\bea
  (\psi^{\otimes\, n})^I_J  = \psi^{i_1}_{j_1}\psi^{i_2}_{j_2}\cdots \psi^{i_n}_{j_n}\qquad
  (\psi^{\dagger \otimes\, n})^K_L  = \psi^{\dagger\, k_n}_{l_n}\cdots \psi^{\dagger\, k_2}_{l_2} \psi^{\dagger\, k_1}_{l_1}
  \label{daggerorder}
\eea
for ordering, the two point function is given by
\bea
  \langle (\psi^{\otimes\, n})^I_J(\psi^{\dagger \otimes\, n})^K_L\rangle 
= \sum_{\sigma\in S_n} {\rm sgn}(\sigma)\sigma^I_L(\sigma^{-1})^K_J
\label{wickforfermions}
\eea
where ${\rm sgn }(\sigma)$ is the sign of permutation $\sigma$. 
The sign of the permutation is given by ${\rm sgn}(\sigma)=(-1)^m$ where $m$ is the number of transpositions in the product.
The ordering in (\ref{daggerorder}) is adopted to ensure that there are no $n$ dependent phases in (\ref{wickforfermions}).

Based on experience with the bosonic case, we expect the Schur polynomials are a linear combination of traces
\bea
 \sum_{\sigma\in S_n}C_\sigma {\rm Tr}_{V^{\otimes n}}(\sigma \psi^{\otimes \, n})
\eea
The anti-commuting nature of the fields must be reflected in the above sum.
To see how this happens, consider changing summation variable from $\sigma$ to $\gamma^{-1}\sigma\gamma$.
The permutation $\gamma$ swaps fields inside the trace. Since we are swapping fermions, we get $-1$ for each swap so that%
\bea
 \sum_{\sigma\in S_n}C_\sigma {\rm Tr}_{V^{\otimes n}}(\sigma \psi^{\otimes \, n})
&=& \sum_{\sigma\in S_n}C_{\gamma^{-1}\sigma\gamma} {\rm Tr}_{V^{\otimes n}}(\sigma \gamma \psi^{\otimes \, n}\gamma^{-1})\cr
&=& \sum_{\sigma\in S_n}C_{\gamma^{-1}\sigma\gamma} {\rm sgn}(\gamma){\rm Tr}_{V^{\otimes n}}(\sigma \psi^{\otimes \, n} )
\eea
Swapping fields must be a symmetry of the basis so that the coefficients must obey
\bea
  C_{\gamma^{-1}\sigma\gamma}={\rm sgn}(\gamma) C_\sigma
  \label{keyproperty}
\eea
This is enough to determine the coefficients $C_\sigma$.
To show this we will make use of the Clebsch-Gordan coefficient for $R \times R$ to couple to the antisymmetric irrep 
$[1^n ]$, denoted by $S^{[1^n]\,R\, R}_{\quad\,\, m\, m'}$.
These Clebsch-Gordan coefficients obey (this is obtained by specializing formula 7-186 of \cite{Hammermesh})
\bea
  \Gamma^{R}_{ij}(\sigma)\Gamma^{R}_{kl}(\sigma)S^{[1^n]\,R\, R}_{\quad\,\, j\, l}={\rm sgn}(\sigma) S^{[1^n]\,R\, R}_{\quad\,\, i\, k}
\eea
Assume without loss of generality that we have an orthogonal representation, so that
\bea
  S^{[1^n]\,R\, R}_{\quad\,\, m\, l}\Gamma^{R}_{lk}(\sigma) ={\rm sgn}(\sigma)\Gamma^{R}_{mi}(\sigma)S^{[1^n]\,R\, R}_{\quad\,\, i\, k}\cr\cr
\Rightarrow\qquad  \Gamma^S(\sigma )O={\rm sgn}(\sigma )\, O\Gamma^S(\sigma )
\label{anticom}
\eea
where $O_{m\, m'}=S^{[1^n]\,R\, R}_{\quad\,\, m\, m'}$. 
$OO^T$ commutes with every element of the group and hence is proportional to the identity matrix.
Thus, after suitable normalization, we have

\bea
  O O^T ={\bf 1}
\eea
$O$ can only be non-zero for self conjugate irreps because $S^{[1^n]\,R\, R}_{\quad\,\, m\, m'}$ is only non-zero for self conjugate irreps.
Recall that given a Young diagram, the conjugate (or transposed) diagram is obtained by exchanging the roles of the rows 
and columns. 
A self-conjugate irrep is labeled by a Young diagram which coincides with its conjugate diagram.
Note that
\bea
  C_{\gamma^{-1}\sigma\gamma}&=&{\rm Tr}\left(O\Gamma^R(\gamma^{-1}\sigma\gamma)\right)\cr
&=&{\rm sgn}(\gamma){\rm Tr}\left(\Gamma^R(\gamma^{-1})O\Gamma^R(\sigma)\Gamma^R(\gamma)\right)\cr
&=&{\rm sgn}(\gamma) C_\sigma
\eea
which proves that the coefficients of our polynomials do indeed obey (\ref{keyproperty}).
Using these coefficients we immediately obtain the Schur polynomials for fermions.
Spelling out index structures, our conventions are
\bea
  \chi_R(\psi)={1\over n!}\sum_{\sigma\in S_n} \chi_R^F(\sigma)
                          \psi^{i_1}_{i_{\sigma (1)}}\cdots\psi^{i_n}_{i_{\sigma (n)}}\qquad
  \chi_R^\dagger (\psi)={1\over n!}\sum_{\sigma\in S_n} \chi_R^F(\sigma)
                          \psi^{\dagger\,\,i_n}_{i_{\sigma (n)}}\cdots\psi^{\dagger\,\,i_1}_{i_{\sigma (1)}}
\eea
where we have introduced the {\it twisted character}
\bea
  \chi_R^F(\sigma)\equiv {\rm Tr}\left(O\Gamma^R(\alpha)\right)
\eea
The two point function is easily evaluated, to give
\bea
  \langle \chi_R \chi_S^\dagger\rangle = \delta_{RS}f_R
\eea
where $f_R$ is the product of factors, one for each box, of Young diagram $R$.
Recall that the box in row $i$ and column $j$ has factor $N-i+j$.

\section{Evaluation of $\chi_R^F(\sigma)$ for the hook $R$}\label{twistedchar}

In this section we will evaluate the twisted character when $R$ is the hook representation.
In this case $R$ has a single row of length $>1$ and many rows of length 1.
As we explained in the last section, this twisted character is only non-zero if the representation $R$ is self conjugate.
This implies that the number rows of length 1 in $R$ is equal to the number of columns of length 1.
Our argument uses Young's orthogonal representation for the symmetric group, which is reviewed in Appendix \ref{YOR}.

Before turning to the evaluation it is useful to review the explicit formula given in \cite{Hammermesh} for the
Clebsch-Gordan coefficient $S^{[1^n]\,R\, R}_{\quad\,\, m\, m'}=O_{mm'}$.
This requires that we know something about how to label states in a given symmetric group irrep $R$.
Towards this end, recall that a Young diagram with $n$ boxes can be filled with a unique integer $1,2,\cdots,n$ in each box.
A tableau is called standard if the entries in each row and each column are increasing.
For every standard tableau there is a unique state in the vector space carrying representation $R$ and the dimension
of symmetric group irrep $R$ is given by the number of standard tableau that can be obtained by filling $R$.
For the self conjugate Young diagrams introduce the notation
\bea
|i\rangle =\young(124,35,6)\qquad\qquad\qquad\qquad\qquad |i^T\rangle =\young(136,25,4)
\eea
which implies that there is a natural pairing of the states belonging to a self conjugate irrep.
Consider the standard tableau
\bea
|i=1\rangle =\young(123,45,6)
\eea
Any other standard tableau is given a sign depending on how many swaps are needed to get it to match $|1\rangle$.
For an even number of swaps the sign is $+1$ and for an odd number it is $-1$.
For pattern $i$, denote this sign by $\Lambda_i$.
Then formula (7-211a) of \cite{Hammermesh} says
\bea
O=\sum_{i=1}^{d_R}\Lambda_i |i\rangle\langle i^T|
\eea
with $d_R$ the dimension of irrep $R$.
Evaluating the character is computing the sum
\bea
\chi_R(O\sigma)=\sum_{i=1}^{d_R}\Lambda_i\,\, \langle i^T|\Gamma^R(\sigma )|i\rangle
\eea
Notice that $O$ has no diagonal elements.
There is only a non-zero contribution to the twisted character when  $\sigma$ can turn $|i\rangle$ into $|i^T\rangle$.
In addition, because of (\ref{anticom}) only elements with ${\rm sgn}(\sigma)=1$ can have a non-zero twisted character.
Note also that elements in the same conjugacy class have the same character, up to a sign
\bea
  \chi_R^F(\sigma)&=&{\rm Tr}\left(O\Gamma^R(\alpha)\right)=
{\rm Tr}\left(O\Gamma^R(\rho)\Gamma^R(\rho^{-1})\Gamma^R(\alpha)\right)\cr\cr
&=&{\rm sgn}(\rho){\rm Tr}\left(O\Gamma^R(\rho^{-1})\Gamma^R(\alpha)\Gamma^R(\rho)\right)
={\rm sgn}(\rho)\chi_R^F(\rho^{-1}\sigma\rho)
\eea
Thus, to prove the character of a given permutation vanishes we can study any permutation in the conjugacy class.

With these observations in hand, we will now argue that only a single conjugacy class has a non-vanishing twisted
character when $R$ is a hook.  
If a hook representation is to be self conjugate, the corresponding Young diagram must have an odd number of boxes.
Consider a hook with $2k+1$ boxes. 
Flipping the standard tableau implies that all of the labels in the pattern greater than 1 change position.
From the structure of Young's orthogonal representation we know that a permutation, after it is written in terms of adjacent transpositions, only swaps the labels of boxes that are named in the permutation.
Thus, to satisfy the fact that all labels greater than 1 change position we know that all labels greater than 1 must appear.
A $2k+1$ cycle may give a non-zero result.
One might expect a $2k$ cycle will give a non-zero result.
This is not the case.
We can see this in two ways as follows: (i) the $2k$ cycle is odd so we know its twisted character vanishes and (ii) we could use the $2k$ cycle $(1,2,3,\cdots,2k-1,2k)$ that leaves $2k+1$ inert.
This permutation never moves label $2k+1$ and hence never changes $|i\rangle$ into $|i^T\rangle$.
This second observation shows that we need all $2k+1$ labels to appear in the permutation for a non-zero result.

Now, use all the $2k+1$ labels to form a permutation built from smaller disjoint cycles.
At least one of these cycles must have an odd length.
The labels of boxes can be shuffled between boxes named in a given cycle, but we only mix boxes named in the same cycle. 
Choose the cycle with odd length to have the form $(2k+1,2k,2k-1,...,2i+1,2i,2i-1)$. Its clear that there is no way
to obtain $|i^T\rangle$ from $|i\rangle$ by just shuffling the labels of these boxes.
Consequently we conclude that for the self conjugate hook representation with $2k+1$ boxes, only the $2k+1$ cycle gives a non-zero twisted character.

We will now evaluate the only non-zero character $\chi_R^F(\sigma)$ which is for $\sigma$ a $2k+1$ cycle.
The evaluation is a straightforward application of the rules of Appendix \ref{YOR}.
The important aspects of the computation are the following
\begin{itemize}
\item[1.] The only states that contribute to the character have patterns such that $2i+1$ and $2i$ for $i=1,2,\cdots,k$
appear in different arms (horizontal or vertical) of the hook. This implies that a total of $2^k$ states contribute to the character.
\item[2.] The character picks up a factor of ${\sqrt{(2i+1)(2i-1)}\over 2i}$ for $i=1,2,\cdots,k$, because $2i+1$ and $2i$
are swapped.
\item[3.] The labels $2i$ and $2i-1$ are not swapped. If these labels are in the same arm we pick up a factor of $\pm1$ and if these factors are in different arms we pick up a factor of $\pm {1\over 2i-1}$, for $i=1,2,\cdots,k$.
In the end the signs conspire so that only the overall sign is not fixed.
\end{itemize}
The result for the only non-zero twisted character is
\bea
\chi_R^F(\sigma)&=&\pm \prod_{i=1}^k {\sqrt{(2i+1)(2i-1)}\over 2i}\left( 1+{1\over 2i-1}\right)\cr
&=&\pm\sqrt{2k+1}
\eea
This has an interesting and immediate consequence for the Schur polynomial $\chi_R(\psi)$ when $R$ is a self conjugate
hook.
Recall that
\bea
  \chi_R(\psi)={1\over (2k+1)!}\sum_{\sigma\in S_{2k+1}} \chi_R^F(\sigma)
                          \psi^{i_1}_{i_{\sigma (1)}}\cdots\psi^{i_{2k+1}}_{i_{\sigma (2k+1)}}
\eea
The only contribution to the sum is for $\sigma$ a $2k+1$ cycle, and there are $(2k)!$ such terms.
The sign of the character is correlated with the sign of the trace 
$\psi^{i_1}_{i_{\sigma (1)}}\cdots\psi^{i_{2k+1}}_{i_{\sigma (2k+1)}}=\pm \Tr (\psi^{2k+1})$, so that in the end we find
\bea
  \chi_R(\psi)={1\over \sqrt{2k+1}}\Tr (\psi^{2k+1})\label{HookResult}
\eea
which is a special case of the general result we prove in this paper.

Before concluding this section, we note that given the above value of the twisted character, there is a straightforwards extension to representations $R$ made up by stacking hooks.
As an example, stacking hooks of length 13, 7 and 3 produces
\bea
{\small \yng(7,5,4,3,2,1,1)}
\eea
Label the representation by the hook lengths of the stacked hooks.
The above representation is labeled $(13,7,3)$.
We will evaluate $\chi_R^F(\sigma)$ with $R$ the representation
$(2k_1+1,2k_2+1,\cdots,2k_l+1)$ and $\sigma$ a permutation with cycle structure\footnote{Here $(n)$ denotes a cycle of length $n$. The permutation $(n)(m)$ comprises a disjoint $n$-cycle and $m$-cycle.} $(2k_1+1)(2k_2+1)\cdots(2k_l+1)$.
We will write $\sigma=\sigma_{2k_1+1}\sigma_{2k_2+1}\cdots\sigma_{2k_l+1}$. 
We make 3 basic observations:
\begin{itemize}
\item[1.]Not all states contribute to the trace.
In going from $|i\rangle$ to $|i^T\rangle$ it is clear that boxes are not swapped between hooks.
Thus the labels appearing in the $2k_q+1$-cycle of $\sigma$ must all populate the hook of length $2k_q+1$.
All states that don't obey this condition can be dropped as they don't contribute.
\item[2.] As reviewed in Appendix \ref{YOR}, the action of a given permutation is determined by the content of the Young diagram. 
This content is the same for the stacked or un stacked hooks. 
Thus, the action of the $(2k_l+1)$ cycle on the hook of length $2k_l+1$ is the same whether or not it is stacked in $R$. 
An illustration of this rule for the content is as follows
\bea
\begin{ytableau}
0&1&2&3&4&5\\
-1&0&1&2\\
-2&-1&0\\
-3&-2\\
-4\\
-5
\end{ytableau} \quad 
\begin{ytableau}
0&1&2&3&4&5\\
-1\\
-2\\
-3\\
-4\\
-5
\end{ytableau} \quad 
\begin{ytableau}
0&1&2\\
-1\\
-2\\
\end{ytableau} \quad 
\begin{ytableau}
0
\end{ytableau}
\eea

\item[3.] Consider the full set of states in $R$ that participate to the fermionic character. They can be decomposed into 
labels, one for each hook, and the labels runs over all the states of that hook.
\end{itemize}

The above observations taken together imply that there is a tensor product structure to the subspace of states in $R$ that contribute to the character and further that the action of the permutation $\sigma$ splits up so that each (disjoint) cycle
in $\sigma$ acts on a different hook. 
Thus
\bea
\chi_R^F (\sigma)&=&\sum_I \langle R,I|\sigma_{2k_1+1}\sigma_{2k_2+1}\cdots\sigma_{2k_l+1}|R,I\rangle\cr\cr
&=&\sum_{i_1,i_2,\cdots i_l} \langle (2k_1+1),i_1|\otimes\langle (2k_2+1),i_2|\otimes\cdots\otimes \langle (2k_l+1),i_l|
\sigma_{2k_1+1}\sigma_{2k_2+1}\cdots\sigma_{2k_l+1}\cr
&&\qquad\qquad\qquad |(2k_1+1),i_1\rangle\otimes|(2k_2+1),i_2\rangle\otimes|(2k_l+1),i_l\rangle\cr\cr
&=&\sum_{i_1} \langle (2k_1+1),i_1|\sigma_{2k_1+1}|(2k_1+1),i_1\rangle
\cdots\sum_{i_l} \langle (2k_l+1),i_l|\sigma_{2k_l+1}|(2k_l+1),i_l\rangle\cr\cr
&=&\chi_F^{(2k_1+1)}(\sigma_{2k_1+1})\cdots \chi_F^{(2k_l+1)}(\sigma_{2k_l+1})\cr\cr
&=&\pm\prod_{i=1}^l \sqrt{2k_i+1} \label{eq:charmt}
\eea
where in the last line we used the fermionic character for a hook.

This is almost the proof we need. We still need to show that no other cycle structure has a non-trivial character on a given tableau as above.

\section{Fermion Schur polynomials are traces}\label{proof}

The result obtained in (\ref{HookResult}) shows that the fermion Schur polynomial labeled by a hook is given
by a single trace.
In this section this result will be generalized to any self conjugate representation $R$, where there is a single trace structure that contributes.

Any such representation is obtained by stacking self conjugate hooks.
The result we will prove shows that the Schur polynomial for a representation obtained by stacking $k$ hooks
is equal to a product of $k$ traces, one for each hook.
The number of fields inside the trace equals the number of boxes in the corresponding hook.
$k$ is equal to the number of boxes on the diagonal in $R$.

The expression \eqref{eq:charmt} will give the normalization that relates the trace structure to the tableau.
To finish the proof we employ an induction argument. We use the idea that single traces can be treated as orthogonal particles at large $N$: each trace represents a 
different creation operator, so that when we multiply by a trace we should get a new element of the Fock space of states with that particle present.

We will assume that we have proved the result for all tableaux with at most $k$ self conjugate hooks. 
In the particle language, this is assuming that we have proved the result for all occupation numbers less than or equal to $k$. The one hook 
result is the one particle result. In that case, the single trace is equal to a single hook times the square root of the number of fields in the trace.

The idea is then to add an extra trace (particle) and to prove the result with the extra trace included, which should now have $k+1$ hooks and to account for the new states that are generated this way.
Since we have already shown the result for $k=1$, this will prove the result by induction for all $k$.

Consider a trace structure of the form $\tr(\psi^{s_1}) \dots \tr(\psi^{s_k})$. 
Using cyclicity and the anticommuting nature of $\psi$ it is easy to see that each power $s_i$ $i=1,\cdots,k$ 
is odd for a non-zero trace.
Therefore each trace is an anticommuting variable. 
This means that for a non-zero product all the $s_i$ must be distinct from each other. 
Order the $s_i$ in decreasing order $s_1>s_2>\dots s_k$. 
Also, associate to this trace structure the following permutation in $S_L$  with $L=\sum_{i=1}^k s_i$
\begin{equation}
(1, 2, 3 \dots, s_1) (s_1+1, s_1+2, \dots, s_1+s_2) \dots (s_1+\dots +s_{k-1}+1, \dots, L)
\end{equation}
This permutation is constructed by taking the numbers from $1,\dots, L$ and doing an ordered cycle on the first $s_1$ elements $1, \dots, s_1$, a cycle of order $s_2$ on the next set of elements etc. This gives a unique element of the permutation group for each trace structure.

Now let the above permutation act on a self conjugate representation labeled by a Young diagram with $L$ boxes.
To get a non-zero answer, when acting an a given standard tableau the  labeling must be reflected by the permutation that we have chosen. For example, the action of the permutation must send
\bea
\begin{ytableau}
1&2&4&6\\
3&8&9&11\\
5&10&13\\
7&12
\end{ytableau}
\qquad \longrightarrow\qquad 
\begin{ytableau}
1&3&5&7\\
2&8&10&12\\
4&9&13\\
6&11
\end{ytableau}\label{ExTab}
\eea

The idea now is to add an extra trace $\tr(\psi^{s_{k+1}})$ to the structure, with $s_{k+1}< s_k$.  That is, we want to multiply the fermion Schur polynomial by the single hook tableaux with $s_{k+1}$ elements. 
We will now do induction on $s_{k+1}$ to check what the structure of the product tableaux should be.

The first thing to notice is that a single permutation $(a_1, a_1+1,\dots, a_1+2s)$ can only perform reflections between a pair of consecutive $a+i$, $a+{i+1}$. To see this,  note that  the cycle permutation can be written as
\begin{equation}
(a, a+1,\dots, a+2s)= (a+2s-1, a+2s)\circ\dots\circ (a+1,a+2)\circ (a, a+1)
\end{equation}
and $a+i$ appears in at most two places. These are the only times in which the permutation can move $a+i$. 

Since the number of elements inside the permutation is odd, a reflection of the labels must have a fixed point. Such a fixed point needs to be a fixed point of the labeling of the standard tableau. For the example discussed in (\ref{ExTab}) above,  $1, 8$ and $13$ are fixed points. This means that a product of $k$ traces needs to have at the least $k$ fixed points. Now we will show that it has exactly $k$ such fixed points. 

To prove this consider the representation of $U(N)$ associated to the trace structure we start with and tensor it with the single self conjugate hook of $s_{k+1}= 2n-1$ boxes. Associate to this hook the labeling
\begin{equation}
\begin{ytableau}
1&1&1&\dots & 1\\
2\\
3\\
\vdots
\\
n
\end{ytableau}
\end{equation}
When we tensor a fixed representation $T$ with this one (in the sense of representations of $U(N)$), we need to sum over all tableaux where we have added exactly $2n-1$ boxes to $T$.
We then distribute the labels in the boxes above in the boxes added to $T$ respecting the following rule (the Littlewood-Richardson rule). Read the numbers in reverse order from the right to the left in each successive row. 
This will produce a pattern like $1, 1, 2, 3, 1, \dots $ which we call the semistandard tableau associated to the product.
The rules for filling the semistandard tableau are that in each row, the order of the labels is non-increasing from right to left, and in each column the labels are strictly decreasing from bottom to top (increasing from top to bottom). The second rule (adapted to the present case) is that there is at least one label $1$  before the label $2$, one $2$ before a $3$ etc in the word of the pattern.
One can check that it is impossible to add boxes in a shape of a $2\times 2$ square, while respecting these rules. 
This means that the result is to take the tableaux and add a collection of skew-hooks to it that are not touching, with the understanding that corners don't count.
For example
\begin{eqnarray}
\begin{ytableau}
\ &\ \\
\
\end{ytableau}\otimes \begin{ytableau}
1 &1 \\
2
\end{ytableau} &=& \begin{ytableau}
\ & \  & 1& 1\\ 
\ & 2
\end{ytableau}+  \begin{ytableau}
\ & \  & 1\\ 
\ & 1& 2
\end{ytableau}\nonumber\\
&+&\begin{ytableau}
\ & \  & 1\\ 
\ & 1\\
2
\end{ytableau}+\begin{ytableau}
\ & \  & 1\\ 
\ & 2\\
1
\end{ytableau}\nonumber\\
&+& \begin{ytableau}
\ & \ \\ 
\ & 1\\
1&2
\end{ytableau}+
\begin{ytableau}
\ & \ \\ 
\ & 1\\
1\\
2
\end{ytableau}
\end{eqnarray}
In the example illustrated above, only the terms in the middle line of the RHS would give a self-dual diagram. We have just described one way to compute Littlewood Richardson coefficients: we count the number of words with tableaux that have fixed shape in the product and count the patterns that satisfy the rule. These coefficients are notoriously hard to compute, and some asymptotic results are known for large Young tableaux (see \cite{Pak} and references therein).

The point of this discussion is that since we can not add a $2\times 2$ square, we are adding at most one diagonal box. Moreover, because we are adding an odd number of boxes, for the diagram to be self-dual it must have one fixed box in the  reflection. Hence, we are adding at most one diagonal box and we are also adding at least one diagonal box.
In this way we get that the number of traces is exactly equal to the number of diagonal boxes. 

To show that Schurs and trace structures are as given in the isomorphism, we are going to do induction in the numbers of traces.  We assume that we have proved it for all $k_0\leq k$ and for arbitrarily big tableaux.
Now lets us do induction on the value of $s_{k+1}$ itself. 
Namely, we do induction on the total particle number and a second induction on the energy of the lightest particle.

Let us start with the smallest possible value $s_{k+1}=1$. We can only add one self dual corner, and we can only do it if all the other $s_k$ are higher. For example
\begin{equation}
\begin{ytableau}\ & & \\
\  \\
\
\end{ytableau} \otimes \begin{ytableau}\
\end{ytableau}= \begin{ytableau}\ & & \\
\ & \ \\
\
\end{ytableau}
\end{equation}
is allowed, but
\begin{equation}
\begin{ytableau}\ & & \\
\ & \ \\
\
\end{ytableau} \otimes \begin{ytableau}\
\end{ytableau} = 0
\end{equation}
because we can not add a hook of length one on the diagonal for the second term..

 This procedure shows that we have generated all possible diagrams with exactly one  box in the $k+1$ hook: we had all possible diagrams with $k$ hooks before (by hypothesis)
 and now we have produced all states with $k+1$ hooks where the last hook has length one.
 
 At large $N$ each such selfdual diagram with $M$ boxes has norm $N^M$. Dividing by $N^M$, which only depends on the number of boxes and not the particular shape, we get an orthonormal basis.
  
Let us go to the next case: by multiplying by $\tr(\psi^3)$, due to large $N$ factorization, we should obtain a state that is orthogonal to all configurations that do not have a $\tr(\psi^3)$ in it. In particular, when we multiply it by previous known diagrams with 
$k$ self-dual hooks, if the trace splits into more than one skew hook in the product, it generates tableaux that have exactly one box in the last diagonal. But we already generated all of these by states that have a $\tr(\psi)$, and by orthogonality of the tableaux states, the coefficient with which they are generated must be zero. 
Hence the three boxes of $\tr(\psi^3)$ must all lie in the same hook.
That is, when we multiply by $\tr(\psi^3)$ the only option is not to divide the skew hook and add to add it using the prescription implied by the equality of the Schur and trace basis. We do the same for $\tr(\psi^5)$: it is easy to show that we can not split the hook into various skew hooks because they lead to states that are already accounted for by states produced with $\tr(\psi)$ or $\tr(\psi^3)$. The induction then becomes straightforward: any new product must be orthogonal to all states where the last hook has lower length, which were already all generated. That is, the last hook is not divided and all the boxes of the hook we are multiplying by must belong to the same hook.

Now, this actually completes the proof that the bases are proportional to each other. The argument uses orthogonality at infinite $N$, but what we are studying is the ring structure of multiplying by traces which is independent of $N$. What we are seeing is how multiplying by traces produces new states. 

The final check is that the norm  of a product of traces is the product of the norms of the individual traces, but this is already implicit in \eqref{eq:charmt}, where we computed the character and showed that we have a product structure.

A neat corollary of the result here is that when we take products of twisted Schurs, the twisted Littlewood Richardson coefficients are all $\pm 1, 0$ (taking into account that the
sign depends on the order in which the product is taken)  and that this is straightforward to determine.

\section{Conclusion}\label{conclusions}

In this paper we have exhibited a remarkable connection between the basis of traces and the basis of Schur functions,
constructed for a single fermionic gauged matrix model: they are the same bases, albeit with different normalizations.
To prove the equivalence we have had to develop some new formulas for twisted characters. The results of this paper
provide a complete set of twisted character values for any permutation and any representation, which is more than
what is known explicitly for the usual symmetric group characters. 
The proof of the relation between the Schur and trace bases itself is performed by doing a double
induction, both on the number of traces and in the number of fields in the last trace added.
It uses representation theory of both the symmetric and unitary group, large $N$ factorization and the ring structure of multiplying by traces and it needs all of these ingredients to work.
The twisted Schurs form a ring and the structure constants of this ring are the twisted Littlewood Richardson coefficients.  
Our results prove that the twisted Littlewood Richardson coefficients are all $\pm 1, 0$.

An important implication of our result is that there is a hidden simplicity that was not previously appreciated, that is,
that there is no mixing of traces in the fermionic model.
Indeed, at large $N$ we expect that expectation values of products of traces factorize, but there should be mixing  corrections of order $1/N^2$. 
For the fermionic matrix model we have studied here, our results prove that there is a similar factorization, but in this case the factorization is exact.
There are still corrections to the norm of individual states that are powers of $1/N^2$, but the system has a well defined notion of particle number at any value of $N$.
It would be fascinating to properly explore the consequences of this factorization.

Our results probably have immediate application to some operator-state problems in CFT.
As an example, there are single fermion sectors in the free ${\cal N}=4 $ SYM in the $SU(1|1)$ sector.
Another interesting extension would be to consider cyclic quivers of both fermions and bosons.

We have also noted that strongly coupled electron systems can sometimes display edge states that carry a $c=1/2$ 
chiral Majorana fermion excitation. It would be interesting to explore this possible connection between the fermionic matrix model and topological phases of matter.

{\vskip 0.5cm}

\noindent
\begin{centerline} 
{\bf Acknowledgements}
\end{centerline} 

We would like to thank Sanjaye Ramgoolam for useful discussions.
The work of D.B. is supported in part by the Department of Energy under grant {DE-SC} 0011702. 
The work of RdMK is supported by the South African Research Chairs
Initiative of the Department of Science and Technology and National Research Foundation
as well as funds received from the National Institute for Theoretical Physics (NITheP).

\appendix

\section{Young's Orthogonal Representation}\label{YOR}

This representation is specified by giving the action of the ``adjacent transpositions'' which are swaps of the form $(i,i+1)$.
A box in row $i$ and column $j$ has content $j-i$. 
Here is an example of a Young diagram with the content of each box displayed
\bea
\begin{ytableau}
0&1&2&3&4&5\\
-1&0&1&2\\
-2&-1&0\\
-3&-2\\
-4\\
-5
\end{ytableau}
\eea
Let the box labeled $a$ in the standard tableau have content $c_a$. 
The state $|ST(a\leftrightarrow a+1)\rangle$ is labeled by the tableau obtained by swapping $a$ and $a+1$ in $|ST\rangle$.
Young's orthogonal representation is defined by
\bea
   (i,i+1)|ST\rangle = {1\over c_{i+1}-c_i}|ST\rangle +\sqrt{1-{1\over (c_{i+1}-c_i)^2}}|ST(a\leftrightarrow a+1)\rangle
\label{DefRep}
\eea
This defines the irrep because any element of the group can be written as a product of adjacent permutations.
The following example is obtained using (\ref{DefRep})
\bea
(3,4) \left|\young(124,35,6)\right\rangle = {1\over 3}\left|\young(124,35,6)\right\rangle
+{\sqrt{8}\over 3}\left|\young(123,45,6)\right\rangle
\eea

\end{document}